\documentclass[12pt,a4paper]{iopart}
\usepackage{times}
\usepackage{graphicx}
\usepackage{cite}

\begin{document}
\title[Non-Stationary Single-Photon Source]{Photon Statistics of a Non-Stationary Periodically
Driven Single-Photon Source}
\author{M. Hennrich, T. Legero, A. Kuhn, and G. Rempe}
\address{Max-Planck-Institut f\"ur Quantenoptik, Hans-Kopfermann-Str. 1, 85748 Garching, Germany}
\ead{axel.kuhn@mpq.mpg.de}
\begin{abstract}
We investigate the photon statistics of a single-photon source that operates under
non-stationary conditions. The photons are emitted by shining a periodic sequence of laser
pulses on single atoms falling randomly through a high-finesse optical cavity. Strong
antibunching is found in the intensity correlation of the emitted light, demonstrating that a
single atom emits photons one-by-one. However, the number of atoms interacting with the cavity
follows a Poissonian statistics so that, on average, no sub-Poissonian photon statistics is
obtained, unless the measurement is conditioned on the presence of single atoms.
\end{abstract}
\pacs{03.67.-a, 03.67.Hk, 42.55.Ye, 42.65.Dr} \submitto{\NJP}

\section{Introduction}
Worldwide, major efforts are made to realise systems for the storage of individual quantum
bits (qubits) and to conditionally couple different qubits for the processing of quantum
information \cite{Bouwmeester00}. Ultra-cold trapped neutral atoms or ions are ideal  quantum
memories that store qubits in long-lived states, while single photons may act as flying qubits
that allow for linear optical quantum computing \cite{Knill01}. On the route to a scalable
quantum-computing network, interconverting these stationary and flying qubits is essential
\cite{DiVincenzo00}.  One way to accomplish such an interface is by an adiabatic coupling
between a single atom and a single photon in an optical cavity \cite{Enk97:1,Cirac97}.

The present work focusses on the properties of a coupled atom-cavity system which is operated
as a single-photon emitter \cite{Kuhn99, Kuhn02:2, Hennrich03, Kuhn03}.   In contrast to other
methods of Fock-state preparation in the microwave regime \cite{Maitre97,Brattke01}, where the
photons remain trapped inside the cavity, our scheme allows one to  emit  single optical
photons on demand into a well-defined mode of the radiation field outside the
cavity\cite{Hennrich00,Kuhn02}. However, in contrast to many other single-photon sources, like
solid-state systems \cite{Kurtsiefer00, Beveratos01, Santori02}, our source operates under
non-stationary conditions, because atoms enter and leave the cavity randomly. Only during the
presence of a single atom, the atom-cavity system is acting as a single-photon emitter. No
photons are generated without atoms, and if more than one atom is present, the number of
simultaneously emitted photons might exceed one. These circumstances have a significant impact
on the photon statistics of the emitted light \cite{Kimble03,Kuhn03:2}, which is analysed here
in detail.

\begin{figure}[t]
   \centering
   \includegraphics[width=0.9\columnwidth]{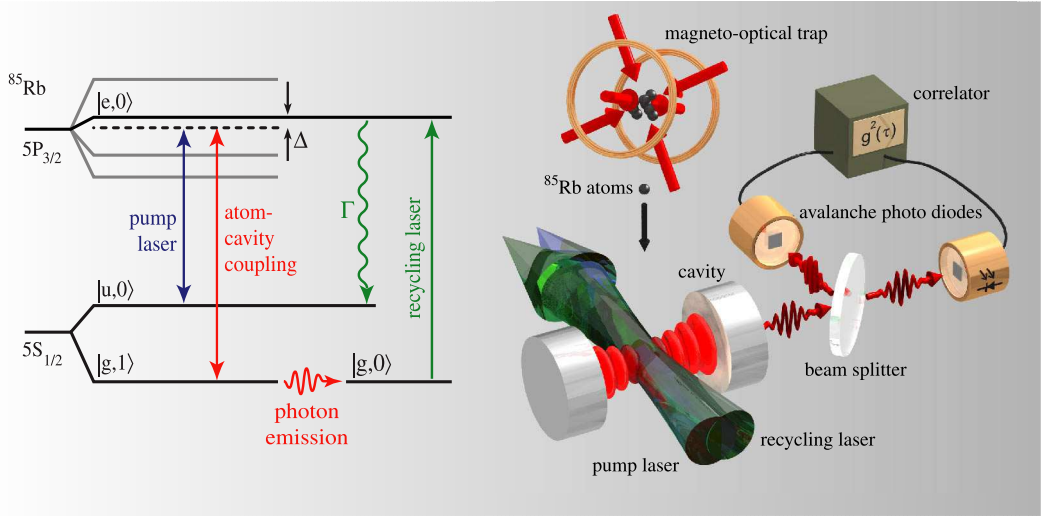}
   \caption{Scheme of the experiment. \textbf{left:}
Relevant levels and transitions in $^{85}$Rb. The atomic states labelled $\left|
u\right\rangle$, $\left| e\right\rangle $ and $\left| g\right\rangle $ are involved in the
Raman process, and the states $\left| 0\right\rangle $ and $\left| 1\right\rangle $ denote the
photon number in the cavity. \textbf{right:} Setup: A cloud of atoms is released from a
magneto-optical trap and falls through a cavity $20\,$cm below in about $8\,$ms with a
velocity of $2\,$m/s. The interaction time of each atom in the cloud with the TEM$_{00}$ mode
of the cavity amounts to about $20\,\mu$s. The pump and recycling lasers are collinear and
overlap with the cavity mode. The light emitted from the cavity is registered by a pair of
photo diodes in Hanbury-Brown \& Twiss configuration in order to analyse the photon
statistics.} \label{setup}
\end{figure}

\section{Single Photons from a Cavity-QED System}
Figure \ref{setup} illustrates the basic scheme of the  process. A dilute cloud of $^{85}$Rb atoms, prepared in state $|u\rangle \equiv |5S_{1/2}(F=3)\rangle$, is released from a magneto-optical trap (MOT) and falls with a velocity of $2\,$m/s through  a 1\,mm long optical cavity of finesse $F=60000$. The density of the cloud, and therefore the average number of atoms simultaneously interacting with the TEM$_{00}$ mode of the cavity,  is freely adjustable. The cavity is near resonant with the transition between  the 
$|5S_{1/2}(F=2)\rangle$ hyperfine state of the electronic ground state and the electronically excited $|5P_{3/2}(F=3)\rangle$ state, labelled $|g\rangle$ and $|e\rangle$, respectively.
Initially, the cavity is empty, so that the state of the coupled atom-cavity system can only move within the Hilbert space spanned by the product states $|u,0\rangle$, $|e,0\rangle$, $|g,1\rangle$ and $|g,0\rangle$, with $|0\rangle$ and $|1\rangle$ denoting the relevant photon number states of the cavity. The dynamics of this system is determined by
$(g_{max},\kappa,\gamma_{\perp},\Delta)/2\pi =  (2.5, 1.25, 3.0,-20.0)\,$MHz, where $g_{max}$ is the cavity-induced coupling between states $|e,0\rangle$ and $|g,1\rangle$ for an atom optimally coupled to the cavity, and $\kappa$ and $\gamma_{\perp}$ are the field and polarisation decay rates of the cavity and the atom, respectively, and $\Delta$ is the detuning of the cavity from the atomic transition. One mirror has a larger transmission coefficient than the other so that photons leave the cavity through this output coupler with a probability of 90\%.   While an atom interacts with the cavity, it experiences a sequence of laser pulses that alternate between triggering single-photon emissions and recycling the atom to state $|u\rangle$: The $2\,\mu$s long pump pulses are detuned by $\Delta$ from the
$|u\rangle\leftrightarrow |e\rangle$ transition, so that they adiabatically drive a stimulated Raman transition (STIRAP)
\cite{Vitanov01,Hennrich03} from $|u,0\rangle$ to $|g,1\rangle$  with a Rabi frequency that increases linearly from 0 to $\Omega_{max}/2\pi=8.0\,$MHz. This Raman transition goes hand-in-hand with a photon emission. Once the photon is emitted, the system reaches $|g,0\rangle $, which is not coupled to the single-excitation manifold, $\{|u,0\rangle, |e,0\rangle, |g,1\rangle\}$, and therefore cannot be re-excited. This limits the number of photons per pump pulse and atom to one.

To emit a sequence of photons from one-and-the-same atom, the system is transferred back to
$\left| u,0\right\rangle $ after each emission. To do so, we apply $2\,\mu$s long recycling
laser pulses that hit the atom between consecutive pump pulses. The recycling pulses are
resonant with the $\left| g\right\rangle \leftrightarrow \left| e\right\rangle $ transition
and excite the atom to state $\left| e\right\rangle$. From there, it decays spontaneously to
the initial state, $\left| u\right\rangle$. In this way, an atom that resides in the cavity
can emit a sequence of single-photon pulses. For each experimental cycle, these photons  are
recorded using  two avalanche photo diodes with 50\% quantum efficiency, which are placed at
the output ports of a beam splitter.

\section{Photon Statistics}
The two photo diodes constitute a Hanbury-Brown \& Twiss setup \cite{Hanbury56} which we use
to measure the normalised intensity correlation of the photon stream emitted from the cavity,
\begin{equation}
g^{(2)}(\tau)=\frac{\langle I_{1}(t) I_{2}(t+\tau)\rangle}{\langle I_{1}(t)\rangle\langle
I_{2}(t)\rangle},
\end{equation}
where $I_{n}(t)$ is the count rate recorded by detector $n=1, 2$. If $\bar I$ denotes the mean
count rate recorded by each detector and $\bar I_{N}$ is the mean noise-count rate, the mean
rate of photon-counts reads $\bar I_{P}=\bar I-\bar I_{N}$. This allows to estimate  two of
the three following contributions to the correlation function:\\
\begin{itemize}
\item[\textbf{(a)}] Correlations between a noise count and either a real photon or another
noise count are randomly distributed and occur with a probability proportional to $\bar
I^2_{N}+\bar I_{P} \bar I_{N}+\bar I_{N} \bar I_{P}=\bar I^2-\bar I_{P}^2$. Therefore these
correlations lead to a constant background contribution,
\begin{equation}
g^{(2)}_{N}=\frac{\bar I^2-\bar I_{P}^2}{\bar I^2}=1-\frac{\bar I_{P}^2}{\bar I^2},
\end{equation}
to the normalised correlation function.

\item[\textbf{(b)}] Correlations between photons that stem from different atoms lead to a
modulation of $g^{(2)}(\tau)$, since the periodicity of the pump laser leads to a modulation
of the photon emission probability. The pump intervals have the same duration as the recycling
intervals, and the probability for photon emissions during recycling is close to zero. This
increases the average rate of photons emitted during pumping to $2 \bar I_{P}$, so that
photon-photon correlations between pump pulses are found with a probability proportional to
$(2\bar I_{P})^2$, while the probability to get photon correlations between pump- and
recycling intervals is vanishingly small. The average normalised contribution of photon-photon
correlations to $g^{(2)}$  therefore oscillates between
\begin{equation}
g^{(2)}_{P,min}=0\qquad\mbox{and}\qquad g^{(2)}_{P,max}=\frac{1}{2} \frac{(2 \bar
I_{P})^2}{\bar I^2}=2 \frac{\bar I_{P}^2}{\bar I^2}.
\end{equation}
In order to obtain this simple estimation,  we use the factor $\frac{1}{2}$ in
$g^{(2)}_{P,max}$ to take into account that photons are emitted only during pump pulses, which
are active half of the time, and we neglect any time-dependence of $I_{P}$ within the pump
pulses. In the experiment, however,
$I_{P}$ varies with time, which causes small deviations from the estimated values, as further
discussed below.
\begin{figure}[t]
   \centering
   \includegraphics[width=0.9\columnwidth]{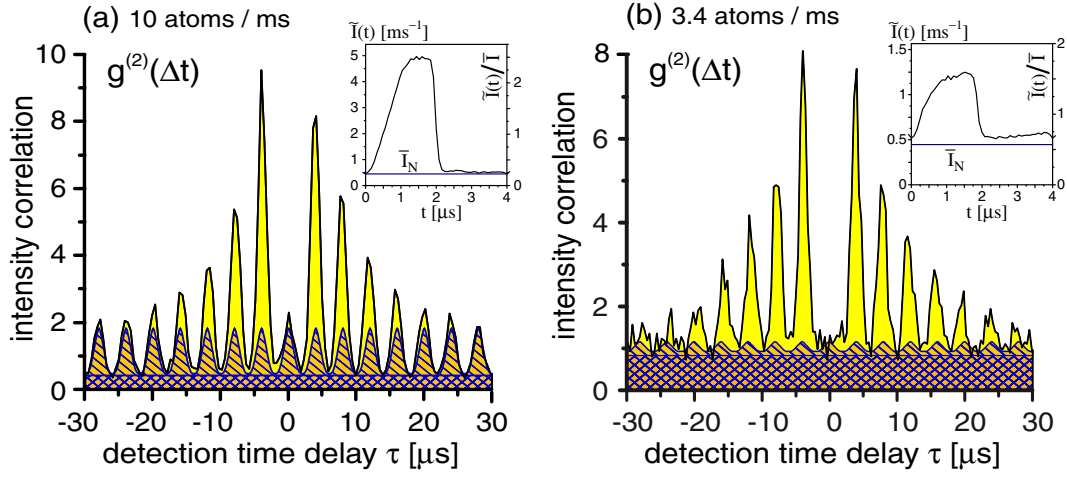}
   \caption{Unconditional photon statistics of the emitted light:  Intensity correlation,
   $g^{(2)}(\Delta t)$, with different-atom (hatched) and noise (cross-hatched) contributions.
   For correlation times larger than the atom-cavity interaction time, only these  contributions
   persist. They are calculated by a convolution of the respective pulse-averaged count rates,
   $\tilde I(t)$, which are shown in the two insets. \textbf{(a)} High atom flux, averaged over
   4997 experimental cycles (loading and releasing of the atom cloud) with a total number of
   151089 photon counts. \textbf{(b)} Low atom flux, averaged over 15000 experimental cycles
   with a total number of 184868 photon counts.}
\label{g2-unconditional}
\end{figure}

\item[\textbf{(c)}] Correlations between photons emitted from one-and-the-same atom are, of
course, most interesting. They cannot be estimated from the average count rates, $\bar I$ and
$\bar I_{N}$. However, due to the limited atom-cavity interaction time, $\tau_{int}$, it is
clear that they only contribute to $g^{(2)}(\tau)$ in the time interval
$[-\tau_{int}\ldots\tau_{int}]$, and therefore this contribution can be distinguished from
\textbf{(a)} and \textbf{(b)} as explained below.
\end{itemize}
Figure \ref{g2-unconditional}(a and b), obtained for a different flux of atoms, shows that the
three contributions above are easily identified in the measured correlation function.  Due to
the limited atom-cavity interaction time, all correlations with $|\tau|\gg \tau_{int}$ either
belong to category \textbf{(a)} or \textbf{(b)}. The oscillatory behaviour of $g^{(2)}(\tau)$
in this regime stems from photons emitted by different atoms, whereas the time-independent
pedestal is mainly caused by correlations involving noise counts. These two contributions are
indicated as hatched and cross-hatched areas, respectively. They were calculated by a
convolution of the pulse-averaged count rate, $\tilde I(t)=\frac{1}{N}\sum_{n=1}^N I(t+n
\tau_{period})$, with a total number of $N$ recorded pump\&recycle intervals of duration
$\tau_{period}=4\,\mu$s. This gives an oscillation of the intensity correlation function
between $g^{(2)}_{C, min}$ and $g^{(2)}_{C, max}$.

These two values of $g^{(2)}$ can also be estimated from the mean count rates, $\bar I$ and
$\bar I_{N}$. This estimation predicts an oscillation of $g^{(2)}$ with the periodicity of the
applied sequence between the two extrema $g^{(2)}_{E, min}=1-\frac{\bar I_{P}^2}{\bar I^2}$
and $g^{(2)}_{E, max}=1+\frac{\bar I_{P}^2}{\bar I^2}$. For the data underlying
Fig.\,\ref{g2-unconditional}, we obtain the following result:

\vspace{0.2cm}
\begin{center}
\begin{tabular}{r|c|c}
&high atom flux, Fig.\,\ref{g2-unconditional}(a) & low atom flux, Fig.\,\ref{g2-unconditional}(b)\\
\hline
$\bar I$&1976\,s$^{-1}$&783\,s$^{-1}$\\
$\bar I_{N}$&446\,s$^{-1}$&446\,s$^{-1}$\\
$\bar I_{P}=\bar I-\bar I_{N}$&1530\,s$^{-1}$&337\,s$^{-1}$\\
\hline
$g^{(2)}_{E, min}\ldots g^{(2)}_{E, max}$&$0.40\ldots 1.60$&$0.81\ldots 1.19$\\
$g^{(2)}_{C, min}\ldots g^{(2)}_{C, max}$&$0.46\ldots 1.82$&$0.91\ldots 1.13$\\
\hline
\end{tabular}
\end{center}
\vspace{0.5cm}

\noindent As mentioned above, the estimated values deviate slightly from the values obtained
from the convolution of the pulse-averaged count rate. This was expected due to the simplified
model our estimation is based on. Note that contributions \textbf{(a)} and \textbf{(b)}
persist also in the regime  $|\tau| < \tau_{int}$, since the atoms have a Poissonian
distribution. Obviously, the excess signal observed here belongs to category \textbf{(c)},
i.e. it reflects the single-atom contribution to the correlation signal. Most remarkably, no
excess signal is found around $\tau=0$, i.e.\ all correlations registered during
one-and-the-same pump pulse either involve noise counts or photons from different atoms.
Correlations between  photons that stem from one-and-the-same atom \textbf{(c)} are only found
between different pump pulses. Moreover, antibunching with $g^{(2)}(0)<g^{(2)}(\tau=n
\tau_{period})$, $n=\pm 1, \pm 2, \ldots$, is observed. This effect cannot be observed for a
classical light source, where the  Cauchy-Schwartz inequality predicts that $g^{(2)}(0)\ge
g^{(2)}(\tau)$ \cite{Walls94}. Therefore the observation of antibunching indicates that a
single atom emits photons one-by-one.

\begin{figure}[b]
   \centering
   \includegraphics[width=0.95\columnwidth]{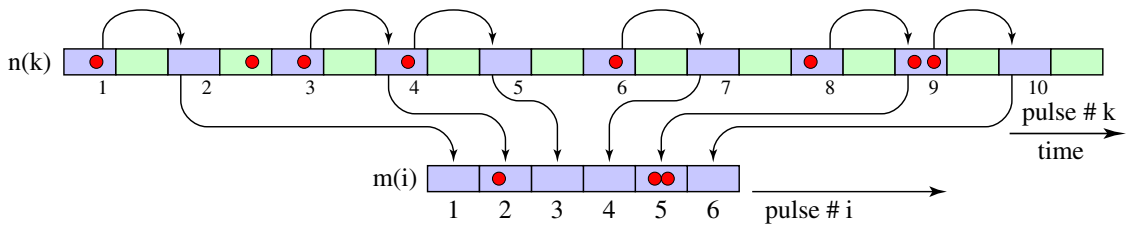}
   \caption{Conditioning on the presence of atoms (schematic): All events (red bullets) are
   first recorded as a function of time. If photons are detected in a time interval where the
   pump laser is active (blue intervals), we assume that an atom is present and take the
   following pump-laser interval into account. Events recorded during recycling (green intervals)
   are ignored. The pump-laser intervals selected this way form a new set of data, which is then
   used to calculate the intensity correlation function.}
\label{conditioning}
\end{figure}

\begin{figure}
   \centering
   \includegraphics[width=0.9\columnwidth]{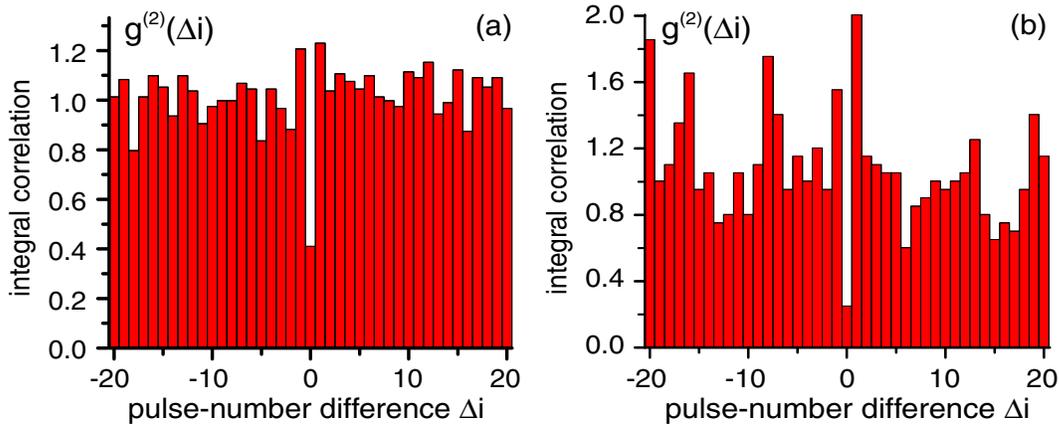}
   \caption{Conditional photon statistics:  A sub-Poissonian photon statistics is found in the
   pulse-to-pulse photon correlation, $g^{(2)}{(\Delta i)}$, conditioned on the presence of
   atoms in the cavity. Data are shown for high atom flux \textbf{(a)} and low atom flux
   \textbf{(b)} as stated in Fig.\,\ref{g2-unconditional}.}
\label{g2-conditional}
\end{figure}

In case of a stationary single-photon source, the non-classicality of the emitted radiation
would lead to a sub-Poissonian photon statistics with $g^{(2)}(0)<1$. In the present case,
however, atoms arrive randomly in the cavity, and the Poissonian atom statistics leads to
$g^{(2)}(0)>1$. An a-priori knowledge on the presence of an atom is therefore needed to
operate the apparatus as a single-photon emitter. Indeed, if the statistical analysis of the
emitted photon stream is restricted  to time intervals where the presence of an atom in the
cavity is assured with very high probability, a sub-Poissonian photon statistics is found.
Figure \ref{conditioning} illustrates the conditioning scheme: To detect an atom, we rely on
the fact that photons are only emitted while an atom resides in the cavity. Thus a photon that
is detected during a pump pulse signals the presence of an atom with probability
$p_{atom}=\bar n_{P}/(\bar n_{P}+\bar n_{N})$, where $\bar n_{N}=2 \bar I_{N} \times \tau_{P}$
is the mean number of noise counts per pump pulse counted by both detectors, and  $\bar
n_{P}=2 \int_{0}^{\tau_{P}}(\tilde I(t)-\bar I_{N}) dt$ is the mean number of detected photons
per pulse, with $\tau_{P}=2\,\mu$s being the pulse duration. For the data underlying
Fig.\,\ref{g2-unconditional}, we obtain $(\bar n_{P}, \bar n_{N}, p_{atom})=(11.6\times
10^{-3}, 1.8\times 10^{-3}, 87\%)$ and $(2.3\times 10^{-3}, 1.8\times 10^{-3}, 56\%)$ for high
and low atom flux, respectively. The small value of $\bar n_{P}$ is due to the fact that the
cavity contains no atom most of the time. But once an atom is detected, it moves only 1/5 of
the cavity waist until the next pump pulse arrives. We can therefore safely assume that the
atom still resides in the cavity at that moment. In the statistical analysis of the light
emitted from the system, we now include only the time interval corresponding to this next pump
pulse. This is accomplished by first denoting the numbers, $k_{1}\ldots k_{M}$, of the $M$
pump intervals with $n(k_{i})>0$, where $n(k_{i})$ denotes the number of photons detected in
the $k_{i}$th interval. The time intervals of the adjacent pump pulses then form the new
stream of selected data, where the number of photons counted with detector 1 and 2,
respectively, is given by $m_{1,2}(i)=n_{1,2}(k_{i}+1)$. From this, the conditioned
correlation function,
\begin{equation}
g^{(2)}(\Delta i)=\frac{1}{M}\sum_{i=1}^M \frac{m_{1}(i)m_{2}(i+\Delta
i)}{\bar{m}_{1}\bar{m}_{2}},
\end{equation}
shown in Fig. \ref{g2-conditional}, is calculated. In case of a small average atom number, see
Fig.\,\ref{g2-conditional}(b), conditioning  yields $g^{(2)}(\Delta i=0)=0.25(11)$, which is
well below one. If the atom flux is  increased, Fig.\,\ref{g2-conditional}(a), one obtains
$g^{(2)}(\Delta i=0)=0.41(6)$. Obviously the larger value is due to the fact that the
probability of having more than one atom interacting with the cavity is not negligible.
Nevertheless, the photon statistics conditioned on the presence of atoms is sub-Poissonian in
both cases, which is demonstrating a noise reduction below the shot-noise level. 
Note that the errors in $g^{(2)}(\Delta i)$ are derived from the standard deviations, $\sigma_{\Delta i}=\sqrt{n_{e}(\Delta i)}$, where $n_{e}(\Delta i)$ denotes the number of events that constitute $g^{(2)}(\Delta i)$ prior to normalisation.  We count $n_{e}(\Delta i=0)=5\pm 2.2$ and $53\pm 7.3$ events, whereas for $\Delta i\neq 0$, we count, on average, $\bar n_{e}=21$ and 130 events with $\sigma_{\Delta i\neq 0}=4.6$ and 11.4, which yields $\bar g^{(2)}(\Delta i\neq 0)=1.00(20)$ and 1.00(9),  for low and high atom flux, respectively.  The fluctuations of $g^{(2)}(\Delta i\neq 0)$ in Fig.\,\ref{g2-conditional} are well explained by this shot noise.

\begin{figure}[t]
   \centering
   \includegraphics[width=0.9\columnwidth]{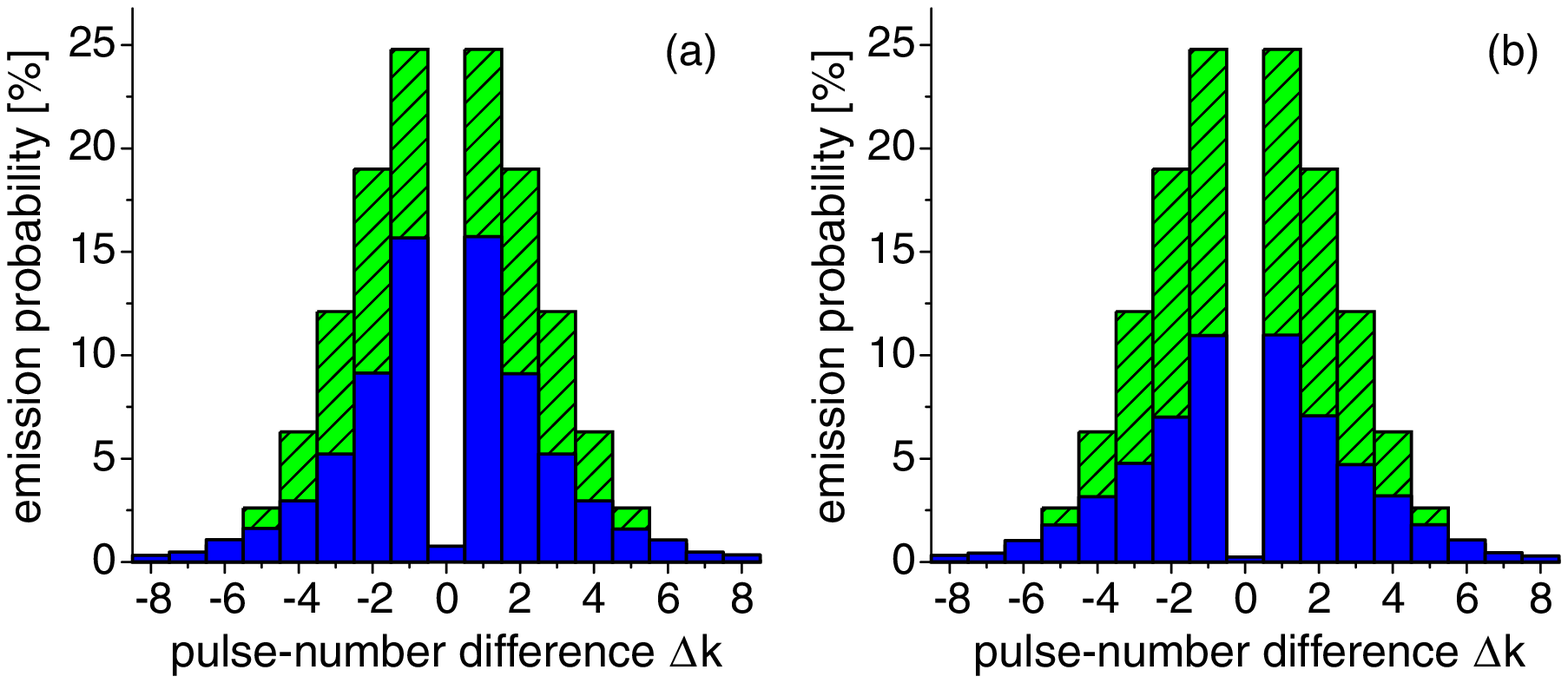}
   \caption{Photon emission probability, conditioned on a photon detection at $\Delta k=0$. The
   theoretical expectation values are represented by the green dashed bars, while the
   experimentally found probabilities are indicated by the solid blue bars. Data are shown
   for high atom flux \textbf{(a)} and low atom flux \textbf{(b)} as stated in
   Fig.\,\ref{g2-unconditional}. The decrease of the experimentally observed probabilities in
   \textbf{(b)} as compared to \textbf{(a)} is attributed to a small misalignment of the setup
   for low atom flux.}
\label{following-pulses}
\end{figure}

From the recorded stream of events, we can also characterise the efficiency of the photon
source by evaluating the probability for the conditional emission of photons during the pump
pulses that follow the detection of an atom by a first photodetection
(Fig.\,\ref{following-pulses}). The evaluation must encompass a correction for detector
effects like noise counts and reduced quantum efficiencies. Hence, we take into account that a
single photodetection actually signals the presence of an atom only with probability
$p_{atom}$, and we also consider that the following photons are detected with an overall
quantum efficiency of $\eta=0.36$ (photo diodes and spatial filtering). We then use the $M$
pump intervals with $n(k_{i})>0$ as starting points to calculate the average photo-emission
probabilities during the neighbouring pulses,
\begin{equation}
\bar p(\Delta k)=\frac{1}{\eta\ p_{atom}}\frac{1}{M}\sum_{i=1}^M \left[n(k_{i}+\Delta k)-\bar
n_{N}-\bar n_{P}-\delta_{\Delta k, 0}\right]. \label{pdelta}
\end{equation}
Here, $n(k)=n_{1}(k)+n_{2}(k)$ is the total number of events counted by both detectors during
the $k$th pump pulse. The mean number of noise counts per pulse, $\bar n_{N}$, and the mean
number of photons per pulse, $\bar n_{P}$, are subtracted from each count number to ensure
that only photons emitted from one-and-the same atom are considered. We also correct for false
triggers, i.e. noise counts signalling atoms which are not present, and the quantum efficiency
of the photodetection. Note that the $M$ conditioning photodetections are not counted twice,
since we subtract $\delta_{\Delta k, 0}$ from the calculated probabilities, with $\delta_{i,
j}=1$ for $i=j$ and $\delta_{i, j}=0$ otherwise. Obviously, the probabilities for subsequent
photon emissions decrease from pulse-to-pulse, since the efficiency of the photon generation
depends on the location of the moving atom. It is highest in an anti-node on the cavity axis
and decreases if the atom moves away from this point. A simulation of the process, based on a
numerical solution of the master equation, allows the calculation of the expected
photon-emission probabilities averaged over the random trajectories of the atoms travelling
through the cavity. This leads to the same qualitative results, but the experimentally
determined emission probabilities are smaller than the expected ones. We attribute this
discrepancy to the random distribution of the atom among its magnetic sublevels after
recycling, which reduces the overall efficiency of the photon generation. In our numerical
simulation, this has been neglected. A more rigourous analysis is beyond the scope of this
paper. Another significant feature of the numerical analysis is the prediction of a
single-photon generation efficiency of $61.6\%$ for an atom which is optimally coupled to the
cavity. Therefore we expect that the present scheme is able to produce single photons in a
highly efficient way provided the atom is hold at rest by, e.g., a dipole-force trap
\cite{Kuhr01,Sauer03,McKeever03:2}.

\section{Summary}
We have statistically analysed the photon stream emitted from a strongly coupled atom-cavity
system in response to laser pulses that adiabatically drive Raman transitions between two
atomic states. The laser  pulses excite one branch of the transition, while  the vacuum field
of the cavity stimulates the other branch. The cavity operates as a non-stationary
single-photon source, since the atoms enter and leave the mode volume randomly, with a maximum
number of about 7 successive photon emissions per atom. Without any a-priori knowledge on the
state of the system, antibunching is observed, which indicates that a single atom emits
photons one-by-one. Furthermore, a preselection has been applied to restrict the analysis to
time intervals where the presence of an atom is assured. This gives a sub-Poissonian photon
statistics with $g^{(2)}(0)<1$. Our setup therefore operates as a deterministic single-photon
emitter, although  the atom statistics is Poissonian.

\section*{Acknowledgments}
This work was partially supported by the  focused research program ``Quantum Information
Processing'' and the SFB631 of the Deutsche Forschungsgemeinschaft, and by the European Union
through the IST(QGATES) and IHP(QUEST and CONQUEST) programs.

\section*{References}

\end{document}